\documentclass[a4,aps,prl,amsmath,twocolumn,showpacs,groupedaddress,floatfix]{revtex4}

\usepackage{graphicx}
\usepackage{caption}

\begin{document}

\title{Imprinting bias stress in functional composites}

\author{V.~Khovaylo}
\affiliation{Institute of Radioengineering and Electronics of RAS,
Moscow 125009, Russia} \affiliation{Physics Department, Moscow
State Mining University, Moscow 119991, Russia}

\author{V.~Koledov}
\affiliation{Institute of Radioengineering and Electronics of RAS,
Moscow 125009, Russia}

\author{E.~Perov}
\affiliation{Institute of Radioengineering and Electronics of RAS,
Moscow 125009, Russia}

\author{V.~Shavrov}
\affiliation{Institute of Radioengineering and Electronics of RAS,
Moscow 125009, Russia}

\author{G.~Lebedev}
\affiliation{State Technological University "Moscow Institute for
Steel and Alloys", Moscow 119991, Russia}

\author{D.~Zakharov}
\affiliation{State Technological University "Moscow Institute for
Steel and Alloys", Moscow 119991, Russia}

\author{M.~Ohtsuka}
\affiliation{Institute of Multidisciplinary Research for Advanced
Materials, Tohoku University, Sendai 980-8577, Japan}

\author{V.~Pushin}
\affiliation{Institute of Metal Physics, Ekaterinburg 620041,
Russia}

\author{H.~Miki}
\affiliation{Institute of Fluid Science, Tohoku University, Sendai
980--8577, Japan}

\author{T.~Takagi}
\affiliation{Institute of Fluid Science, Tohoku University, Sendai
980--8577, Japan}

\begin{abstract}
We propose a simple yet effective method which allows one to
attain large reversible shape changes in shape memory bimetallic
composites without training procedure. It is based on the
conservation of strongly anisotropic martensite microstructure
artificially created in the shape memory layer. This procedure
results in appearance of stress field when the shape memory layer
is transformed to the austenitic state which brings about two-way
shape memory effect. Utilization of this method for preparation of
TiNi-based composite with a thickness of 60 $\mu$m allowed us to
achieve  0.9\% reversible bending deformation. It is also
suggested that the implementation of this method during
preparation of piezoelectric or magnetostrictive composites
permits to imprint bias stress and thus to improve their
characteristics without use of an external load.
\end{abstract}


\date{\today}

\maketitle

The ability of shape memory materials to the reversible
spontaneous shape change -- two-way shape memory effect -- has
widely been used for various applications such as microactuators
or micromachines in microelectromechanical systems (MEMS), various
actuators in electrical appliances, medical implants and
guidewires in biomedical engineering etc. However, this unique
property is not intrinsic to a shape memory material. It is
generally acknowledged that the two-way shape memory effect is due
to the relaxation of residual stresses which promotes formation of
specific martensite variants upon transformation to the
low-temperature state. To create internal stress fields in the
austenitic state, somewhat complicated training procedures such as
introduction of plastic deformation, constraint aging, thermal
cycling, or utilization of precipitates has to be
performed~\cite{1-o,2-z}. This is inconvenient from technological
point of view and increases the total cost of devises. Moreover,
training also results in concomitant effects, e.g. changes of
transformation temperature and hysteresis.

Here we propose a simple yet effective method which allows one to
bring about the two-way shape memory effect in shape memory
bimetallic composites without training procedure. The essence of
the proposed method is that a shape memory layer is bonded with a
metallic layer, as in conventional bimetallic plate, but prior to
the bonding the shape memory layer is stressed in the martensitic
state. Preliminary stressing, particularly tension, creates
strongly anisotropic martensite microstructure which then
conserved by a bonding with the metallic layer.

The appearance of the driving force underlying two-way shape
memory effect in thus prepared composites is schematically
illustrated in Fig.~1. When the shape memory layer is cooled down
below martensite start temperature $M_s$, a specific
microstructure consisting of differently oriented martensitic
variants is formed (Fig.~1a). An external load easily changes this
microstructure, promoting growth of the favorably oriented
martensitic variants and thus leading to seemingly plastic
deformation $\delta l$ (Fig.~1b). If the shape memory layer is
bonded to a metallic layer (Fig.~1c) its strongly anisotropic
microstructure will be conserved. Upon heating above the reverse
martensitic transformation temperature $A_f$ the martensitic
variants disappear and the shape memory layer restores its
original dimensions. Due to the bonding with the metallic layer,
this leads to a strong bending deformation of this bimetallic
composite (Fig.~1d). In the martensite nucleation process
occurring upon cooling below $M_s$ temperature, relaxation of the
bending-induced residual stress field selects certain martensitic
variant and the bimetallic composite reverts its low-temperature
flat shape (Fig.~1e). Since the metallic layer causes recurrent
stress field in the austenitic state of the shape memory layer the
bimetallic composite exhibits two-way shape memory effect.

\begin{figure}[t]
\includegraphics[width=\columnwidth]{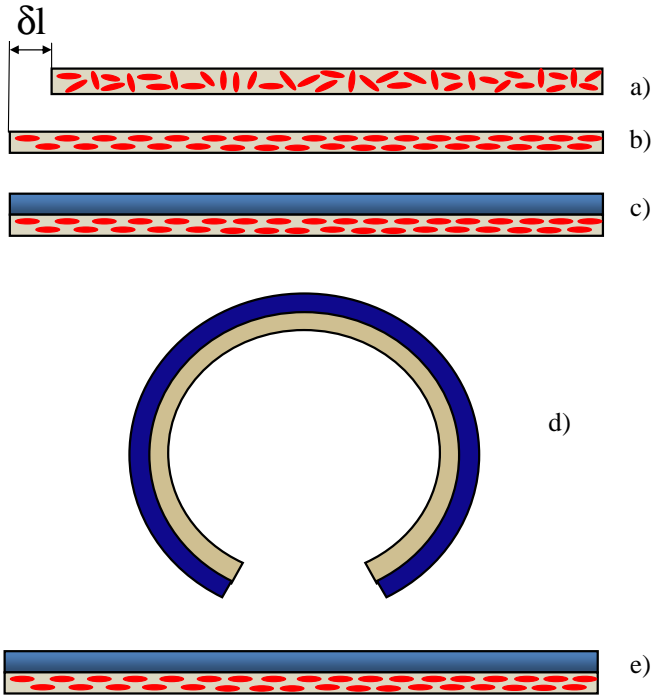}
\caption{Training-free two-way shape memory effect in a bimetallic
composite. Martensitic variants (schematically shown by ellipses)
formed in a shape memory layer after transformation from the
austenitic state (a); Anisotropic martensitic microstructure
created by an external stress (b); Orientation of the martensitic
variants in the shape memory layer imprinted by bonding with a
metallic layer (c); Two-way shape memory effect in the bimetallic
composite: deformation in the austenitic state (d) and the
reversion of the original shape upon transition to the martensitic
state (e).}
\end{figure}

Experimental verification of this method was realized using
Ti$_{50}$Ni$_{25}$Cu$_{25}$ rapidly quenched ribbons ($M_s =
336$~K) with thickness of $30 \mu$m which served as a shape memory
layer. The pseudoplastic deformation  $\delta l \approx  1$\% was
produced in the ribbon by cooling down though the martensitic
transformation temperature under an external load. After that the
Ti$_{50}$Ni$_{25}$Cu$_{25}$ ribbon was glued to a stainless ribbon
of the same thickness. Subsequent experiment showed that the
bimetallic composite exhibits reversible bending/unbending upon
heating/cooling. Similar results were obtained for prepared in the
same manner bimetallic composite made of a
Ni$_{49.5}$Mn$_{28}$Ga$_{22.5}$ thin film ($M_s = 309$~K) and a
stainless ribbon, both of $5 \mu$m thickness.

\begin{figure}[b]
\includegraphics[width=\columnwidth]{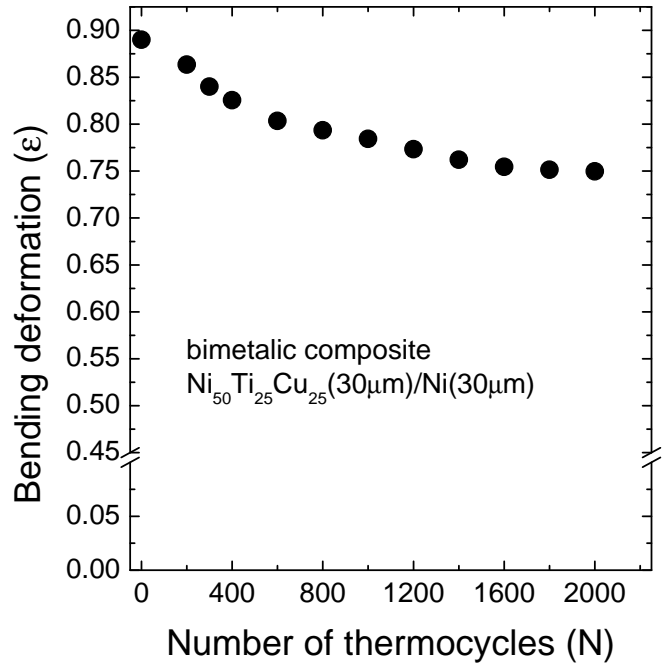}
\caption{Reversible bending deformation   as a function of number
of thermocycles $N$ through the martensitic transition temperature
in a shape memory-based bimetallic composite prepared by
room-temperature electrolytical deposition of a $30~\mu$m Ni layer
onto a Ti$_{50}$Ni$_{25}$Cu$_{25}$ ribbon of the same thickness.}
\end{figure}

Gluing of the constituting layers of the bimetallic composites has
significant shortcoming since their reversible deformation
degrades rather rapidly due to the poor adhesion between the
individual layers. In another composite, a $30~\mu$m Ni layer was
electrolytically deposited at room temperature onto the surface of
Ti$_{50}$Ni$_{25}$Cu$_{25}$ ribbon thermomechanically treated in
the same way as described above. In order to evaluate performance
of the Ti(NiCu)/Ni composite prepared by the electrolytical
deposition, it was subjected to thermocycling through the
martensitic transition temperature. Curvature radius $R$ of the
composite in the austenitic state and the corresponding bending
deformation $\varepsilon  = h/R$ (where $h$ is the thickness of
the composite) has been measured as a function of number of
thermocycles $N$. The results of this experiment (Fig.~2) have
revealed that the bimetallic composite shows $\varepsilon  \approx
0.9$\% upon the first heating/cooling cycle and exhibits fairly
stable performance at least up to 2000 thermocycles through the
martensitic transition temperature. Observed decrease of with the
increase in the number of cycles $N$ is presumably caused by the
accumulation of inelastic defects in the Ni layer.

It has also to be noted that, at the beginning of the experiment
(up to first 400 thermocycles), the composite did not completely
recover its shape in the martensitic state. This presumably means
that the anisotropic distribution of the martensitic variants
created by the external load (Fig.~1b) was not perfectly conserved
by the Ni layer due to rather a large thickness of the
Ti$_{50}$Ni$_{25}$Cu$_{25}$ ribbon. As a result, at the initial
stage of the experiment the nucleation of randomly oriented
martensitic variants took place in lesser strained areas of the
composite. However, it is evident that characteristics of the
composites can be improved by a proper choice of processing
method, adjustment of the thickness of the constituting layers,
and selection of the metallic layer with suitable Young's modulus
and elastic properties.

It must be emphasized that our approach essentially differs from
those reported so far in the literature (see for review
Refs.~[3-6]). Studies of shape memory thin films deposited onto
various substrates have revealed that they frequently exhibit a
two-way shape memory behavior [7-13]. Observed in these composites
reversible bending deformations originate from stress fields which
are induced by a lattice mismatch between the film and substrate.
In our approach, a stress field in the austenitic state of a shape
memory layer is created by the anisotropic martensite
microstructure; the role of the lattice mismatch is immaterial.
This mechanism allows preparation of composites with bending
deformations comparable to the superelastic strains which can be
as large as  6\% in TiNi thin films~\cite{14-i}. Other advantage
of this method is that there is no strict limitation on the
thickness of the constituting layers.

We suggest that the proposed method of imprinting bias stress can
also be used to enhance characteristics of piezoelectric and
magnetostrictive composites. Indeed, it is well documented that
striction of electro- or magnetostrictive materials is markedly
enhanced when they are subjected to mechanical loading which
provides a direction for preferential orientation of structural
domains~\cite{15-b} or magnetization vector~\cite{16-c}. For thin
enough samples, anisotropic microstructure created by electric or
magnetic field can be easy conserved by a deposition of a metallic
layer. For this aim the deposition should be done in the presence
of the field. This method has an additional advantage for
miniaturization of such devises since no mechanical parts are used
for the application of bias stress.

In conclusion, we have proposed and experimentally verified that
in shape memory-based composites a strongly anisotropic martensite
microstructure can be preserved by a deposition of a metallic
layer. This results in large bending deformations of composites
comparable to the superelastic strain of the shape memory layer.
We also have proposed that this approach is capable to improve
characteristics of piezoelectric and magnetostrictive composites.

\end{document}